\documentstyle[12pt]{article}
\begin{document}

\title{Continuous matter fields in Regge calculus}
\author{V.M.Khatsymovsky \\
 {\em Budker Institute of Nuclear Physics} \\ {\em
 Novosibirsk,
 630090,
 Russia} \\ {\em E-mail address: khatsym@inp.nsk.su}}
\date{}
\maketitle
\begin{abstract}
We find that the continuous matter fields are ill-defined in
Regge calculus in the physical 4D theory since the corresponding
effective action has infinite terms unremovable by the UV
renormalisation procedure. These terms are connected with the singular
nature of the curvature distribution in Regge calculus, namely, with
the presence in d $>$ 2 dimensions of the (d-3)-dimensional simplices
where
the (d-2)-dimensional ones carrying different conical singularities are
meeting. Possible resolution of this difficulty is discretisation of
matter fields in Regge background.
\end{abstract}
\newpage
Regge calculus is a (discrete) minisuperspace theory of the {\it
continuum} spacetime, and correspondingly all other fields living in
this spacetime can be taken as either discrete objects or genuine
continuous ones. In particular, the most practically important case of
the electromagnetic field has been considered on a discrete level in
refs. \cite{Sor,Wein}. In ref. \cite{JN1} the scalar field has been
discretised in the 2D Regge calculus. Example of another kind is the
field of the coordinate frame deformations $\xi_\mu (x)$ considered in
ref. \cite{JN2}. This field is defined at each point of the spacetime
continuum independently, i.e. it is considered as ordinary continuous
field.

If continuous matter field is quantised in Regge
background, the problem of vacuum stability can arise.
Indeed, one can say that gravity field strength is infinite
on some subset of points, the conical singularities. On the
other hand, this subset is of zero measure. These two
competing factors make a'priori unclear whether continuous
field is well-defined. The routine way to check it is to
compute the effective action. This reduces (in the case of
free field considered throughout) to calculating some
functional determinants. In the 2D case such calculation
has been performed on the smooth manifold in ref.
\cite{Pol} for the field of arbitrary spin. In ref.
\cite{Sal} it has been proven to be extendable to the
Regge manifold.

In the present paper we show that the continuous matter
fields are ill-defined on the Regge manifold in the
physical 4D case: their contributions into the effective
action suffer from the infinities unremovable by the
standard procedure of the UV renormalisation. The source of
these are points of intersection of different 2D simplices
carrying the conical singularities, i.e. the links.

Consider the conformal trace anomaly $<T^\mu_\mu>$ for a
given field. In
contrast with the 2D case now we can not restore the
effective action from it as in ref. \cite{Pol},
but would like to use it for probing singularities in the
action as quantity derivable from this action.
Potentially dangerous are the terms in $<T^\mu_\mu>$
squared in curvature requiring a care due to the $\delta$-
function nature of the curvature (such the situation does
not arise in two dimensions because the trace anomaly is
linear in curvature there). First we show that nevertheless
the conical singularity alone does not produce unremovable
infinities (at least there is no indication to that).
Indeed, in some coordinates $x$, $y$, $u$, $v$ conical
singularity can be placed in the plane $u$ = 0, $v$ = 0,
and the curvature tensor made to have the only nonzero
component
\begin{equation}
R_{xyxy} \sim \delta (x) \delta (y).
\end{equation}
The bilinear in curvature invariants $R_{iklm}R^{iklm}$,
$R_{ik}R^{ik}$ and $R^2$ entering $<T^\mu_\mu>$ are
proportional to $$A\delta (x)\delta (y)$$ where $A$ stands
for $\delta (0)\delta (0)$ implied in the sense of some
regularisation. This is just proportional to $AR$ and can
be removed by adding to the action the $R$-term, i.e. by
renormalisation of the gravity constant.

Now pass to the intersection of such singularities on a
link. Let us coordinatise the neighbourhood of the link by
introducing radial distance $r$ in each of the adjacent
4-simplices,
\begin{equation}
ds^2=du^2+dr^2+r^2d\Omega^2(\theta ,\phi).
\end{equation}
Here $u$ is a coordinate on the link; $d\Omega^2(\theta
,\phi)$ reduces to a metric on the unit 2-sphere in each
4-simplex, but globally it differs from that one due to
the conical singularities. The only nonzero curvature
component turns out to be
\begin{equation}
\label{Riman}
R_{\theta\phi\theta\phi}=r^2\left ({1\over 2}\,^{(2)}\!
R-1\right )(\gamma_{\theta\theta}\gamma_{\phi\phi}-
\gamma_{\theta\phi}^2)
\end{equation}
if
\begin{equation}
d\Omega^2(\theta ,\phi)=\gamma_{\theta\theta}d\theta^2
+2\gamma_{\theta\phi}d\theta d\phi +\gamma_{\phi\phi}
d\phi^2
\end{equation}
and $\,^{(2)}\! R$ is the scalar curvature of
$\gamma_{ab}$. Thus $\gamma_{ab}$ is required to ensure
${1\over 2}\,^{(2)}\! R-1$ being certain combination of
$\delta$-functions in order to reproduce Regge geometry
in the neighbourhood of a link. Explicit solution of this
problem is not important for us; knowing eq. (\ref{Riman})
allows us to find the curvature invariants,
\begin{equation}
\label{R}
R=2r^{-2}\left ({1\over 2}\,^{(2)}\! R-1\right )
\end{equation}
and
\begin{equation}
\label{RR}
R_{iklm}R^{iklm}=2R_{ik}R^{ik}=R^2={4\over r^4}
\left ({1\over 2}\,^{(2)}\! R-1\right )^2.
\end{equation}
Taking into account the $\delta$-function nature of
$\,^{(2)}\! R$, we find that there are the terms in
$<T^\mu_\mu>$ of the type
\begin{equation}
\label{sing}
Ar^{-4}\delta (\theta -\theta_0)\delta (\phi -\phi_0)
\end{equation}
with infinite $A$. To cancel this the counterterms of
dimensionality $(length)^{-4}$ are required.The only such
ones in the effective action which give nontrivial
contribution to $<T^\mu_\mu >$ are those bilinear in
Riemannian tensor; corresponding contribution to
$<T^\mu_\mu >$ is $\Delta R$. The expression $A\Delta R$
contains the term (\ref{sing}) indeed; but it also contains
the term
\begin{equation}
{1\over 2}Ar^{-4}\Delta_\gamma\delta (\theta -\theta_0)
\delta (\phi -\phi_0)
\end{equation}
which (with reversed sign) thus arises instead of
(\ref{sing}) after subtracting $A\Delta R$. Here
$\Delta_\gamma$ is the 2D covariant Laplacian for the
metric $d\Omega^2$.

Thus, the curvature bilinears (\ref{RR}) are ill-defined in
the neighbourhood of the links of the 4D Regge manifold.
One might wonder why these bilinears were well-defined
(after renormalisation) in the spacetime with an only
conical singularity which can be considered as two
singularities meeting at a link and being continuation of
each other. The matter is in regularisation. The case of
the only singularity is the most symmetrical one when the
spacetime is decomposed as ${\cal M}\times {\cal R}^2$ where
${\cal M}$ is 2D spacetime with conical singularity which
we regularise independently of other two dimensions ($u$,
$v$ above) in ${\cal R}^2$. In the case of more singularities
sharing a link geometry of the problem only allows to
single out the $u$ and radial coordinate $r$ and regularise
in the angle $\theta$, $\phi$ -dependence.

Let us check whether dangerous contributions from different
curvature bilinears cancel each other or not. Write out the
general expression for the trace anomaly \cite{BD},
\begin{equation}
<T^\mu_\mu>={1\over 2880\pi^2}\left\{aC_{iklm}C^{iklm}
+b\left (R_{ik}R^{ik}-{1\over 3}R^2\right )+c\Delta R+
dR^2\right\},
\end{equation}
and, making use of the values of the curvature invariants
(\ref{R}) and (\ref{RR}), we find for the singular part of
$<T^\mu_\mu>$ that
\begin{equation}
\label{T-mu-mu-sing}
<T^\mu_\mu>_{\rm sing}\sim {1\over r^4}
\left ({1\over 2}\,^{(2)}\! R-1\right )^2(2a+b+6d).
\end{equation}
(the term $\Delta R$ is well-defined as distribution).
It is seen to be nonzero (strictly negative) for all the
practically interesting cases: the electromagnetic field
with $a$ = 13, $b$ = -62, $d$ = 0; the spinor field with
$a$ = -7/4, $b$ = -11/2, $d$ = 0; the scalar field with
$a$ = -1, $b$ = -1, $d$ = $-90\left (\xi -{1\over 6}
\right )^2$, $\xi$ being the curvature-scalar coupling.

Note that (\ref{T-mu-mu-sing}) gives only anomalous
contribution. If the field is not conformally invariant as
in the latter case at $\xi$ $\neq$ 1/6, there are also
nonanomalous terms. Since these terms depend on the choice
of quantum state these can cancel the
$<T^\mu_\mu>_{\rm sing}$ only at the expence of imposing a
kind of a constraint on the state. This ruins the parallel
with the continuous counterpart of the theory where such
the constraint is absent.

Thus, matter fields generally can not be defined on Regge
lattice as the genuine continuous objects. A possible way
out of this difficulty may be discretisation of these.

\bigskip
This work was supported in part by the RFBR grant
No. 00-15-96811.


\begin{thebibliography}{99}
\bibitem{Sor}
 Sorkin R 1975 {\it J.Math.Phys.~}{\bf 16}~2432.
\bibitem{Wein}
 Don Weingarten 1977 {\it J.Math.Phys.~}{\bf 18}~165.
\bibitem{JN1}
 Jevicki A and Ninomiya M 1985 {\it Phys.Lett.~}{\bf 150B}
 ~115.
\bibitem{JN2}
 Jevicki A and Ninomiya M 1986 {\it Phys.Rev.D~}{\bf 33}
 ~1634.
\bibitem{Pol}
 Polyakov A M 1981 {\it Phys.Lett.~}{\bf 103B}~207, 211.
\bibitem{Sal}
 Aurell E and Salomonson R 1994 {\it Comm.Math.Phys.~}{\bf
 165}~233; Further results on Functional Determinants of
 Laplacians in Simplicial Complexes, 1994, hep-th/9405140.
\bibitem{BD}
 Birrell N D and Davies P C W, Quantum Fields in Curved
 Space, Cambridge, 1982.
\end{thebibliography}
\end{document}